\documentclass[aps,twocolumn,superscriptaddress,showpacs,amssymb,prx,floatfix,longbibliography]{revtex4-2}
\usepackage{graphicx}
\usepackage[]{hyperref}
\hypersetup{colorlinks=true,linkcolor=blue,citecolor=blue,urlcolor=blue,pdfpagemode=UseNone}
\usepackage{amsmath}
\usepackage{amssymb}
\usepackage{bm}
\usepackage{xcolor}

\newcommand{\slrr}      {$T_1^{-1}$}
\newcommand{\tmvo}      {TmVO$_4$}

\DeclareMathOperator{\sech}{sech}

\begin{document}

\title{Spin-echo and quantum versus classical critical fluctuations in TmVO$_4$}

\author{Y-H. Nian}
\affiliation{Department of Physics and Astronomy, University of California Davis, Davis, CA }
\author{I. Vinograd}
\affiliation{Department of Physics and Astronomy, University of California Davis, Davis, CA }
\author{T. Green}
\affiliation{Department of Physics and Astronomy, University of California Davis, Davis, CA }
\author{C. Chaffey}
\affiliation{Department of Physics and Astronomy, University of California Davis, Davis, CA }
\author{P. Massat}
\affiliation{Geballe Laboratory for Advanced Materials and Department of Applied Physics, Stanford University,  CA 94305, USA}
\author{R. R. P. Singh}
\affiliation{Department of Physics and Astronomy, University of California Davis, Davis, CA }
\author{M. P. Zic}
\affiliation{Geballe Laboratory for Advanced Materials and Department of Applied Physics, Stanford University,  CA 94305, USA}
\author{I. R. Fisher}
\affiliation{Geballe Laboratory for Advanced Materials and Department of Applied Physics, Stanford University,  CA 94305, USA}
\author{N. J. Curro}
\email{njcurro@ucdavis.edu}
\affiliation{Department of Physics and Astronomy, University of California Davis, Davis, CA }

\date{\today}

\begin{abstract}

\textbf{Using spin-echo Nuclear Magnetic Resonance in the model Transverse-Field Ising system TmVO$_4$, we show that low frequency quantum fluctuations at the quantum critical point have a very different effect on $^{51}$V nuclear-spins than classical low-frequency noise or fluctuations that arise at a finite temperature critical point. Spin-echos filter out the low frequency classical noise but not the quantum fluctuations. This allows us to directly visualize the quantum critical fan and demonstrate the persistence of quantum fluctuations at the critical coupling strength in TmVO$_4$ to high temperatures in an experiment that remains transparent to finite temperature classical phase transitions. These results show that while dynamical decoupling schemes can be quite effective in eliminating classical noise in a qubit, a quantum critical environment may lead to rapid entanglement and decoherence.}

\end{abstract}
\maketitle
\newpage

\section*{Introduction}


Unconventional superconductivity tends to emerge in materials in the vicinity of a quantum phase transition \cite{balakirev2003signature,ColemanQCreview,tuson,tusonNature2008,MatsudaB122review2014,Kuo2015,KivelsonNematicQCP2015, Schuberth2016,Paschen2020}, however disentangling competing order parameters and the effects of disorder challenge our ability to discern what interactions or effective Hamiltonians drive the essential physics of these materials.  In order to make progress, it is valuable to investigate paradigmatic systems with parameters that can be well controlled.  A prominent example is LiHoF$_4$, a ferromagnet whose behavior is captured by the transverse field Ising model (TFIM) \cite{RosenbaumQCmodel}. Recently, TmVO$_4$ has emerged as another model TFIM system, with ferroquadrupolar order  that can be tuned to a quantum phase transition by the application of a magnetic field along the crystalline $c$-axis \cite{Massat2021}. Here, the electric quadrupolar moments of the Tm 4f orbitals couple bilinearly to the lattice strain, giving rise to long-range order through a cooperative Jahn-Teller effect \cite{Gehring1975}. The effective Hamiltonian can be described by coupled quadrupolar moments with $B_{2g}$ symmetry, whereas both $B_{1g}$ strain and $c$-axis magnetic fields couple to the transverse Ising variables \cite{FisherNematicQCP}.

{The distinguishing property of a quantum critical point is that there are large quantum fluctuations of the order parameter
at $T=0$ over all length and time scales \cite{CHN1,CHN2,sachdevbook}. These fluctuations persist to higher temperatures over a range of parameter space, giving rise to a `quantum critical fan' in the phase diagram.
The fluctuations affect bulk properties such as resistivity,  {susceptibility} and specific heat, enabling detailed maps of the quantum critical fan to be inferred from the temperature dependence of these quantities \cite{Ohsugi,YRSnature,StewartHFreview,IshidaPdopedBa122PRB2012,FisherScienceNematic2012}.  The presence of such fluctuations can also be inferred from the temperature dependence of the dynamical susceptibility, which can exhibit $E/T$ scaling in the vicinity of the QCP \cite{Schroder2000,VojtaQCreview,Si2010}.
}

Quantum fluctuations arise due to the presence of competing terms in the Hamiltonian that do no commute (e.g., the transverse field versus the Ising interaction), and their dynamics are driven by the intrinsic properties of the Hamiltonian. Thermal fluctuations, in contrast, are driven by the wide range of states explored in a statistical ensemble at finite temperatures, with low frequency fluctuations that are controlled via extrinsic parameters. Both incoherent thermal and coherent quantum fluctuations contribute to the noise fluctuation spectrum at finite temperature. Away from the quantum critical coupling, however, quantum fluctuations remain at high frequency while classical fluctuations become soft at phase transitions \cite{Frerot2016}.  We demonstrate that this distinction enables us to probe the former in \tmvo\ using nuclear magnetic resonance (NMR) spin echoes while avoiding the latter, thus revealing a clear map of the quantum critical fan in this material.

Several years ago the quantum information community considered the question of how a qubit coupled to a noisy environment undergoes decoherence. To model such behavior, they considered a specific type of coupling to a transverse field Ising model as it is tuned through a quantum phase transition \cite{Quan2006,Gu2010,Chen2013}. This model proved fruitful theoretically, but had never been tested experimentally. In our manuscript we present the first experimental study of this effect. 
For 
TmVO$_4$, where the hyperfine interaction between the nuclear spin and the Ising variables has the form $I_x\sigma_x$, where $I_x$ is the nuclear spin and $\sigma_x$ is the Ising operator corresponding to the transverse field direction, the decoherence rate of the nuclear spins is proportional to the transverse-field fidelity susceptibility.  The non-Kramers doublet ground state of \tmvo\ ensures that the hyperfine interaction has exactly this form, offering a unique opportunity to directly probe the quantum critical fluctuations in the TFIM via their interaction with the $^{51}$V nuclear spins  \cite{Washimiya1970}.

{Distinguishing classical and quantum fluctuations at finite temperatures is a problem of fundamental
interest in many-body physics and quantum information \cite{Grover1,Grover2,Swingle,Sherman}.
Our results demonstrate a key distinction in the effect a quantum critical environment has on decoherence of a qubit compared with classical sources of low frequency noise \cite{Frerot2016,Quan2006,Gu2010,Chen2013}. While classical noise can be filtered out by spin-echo and other dynamical decoupling schemes, coupling to a quantum critical environment leads to a qubit's rapid entanglement and decoherence.

Our findings may also inspire new ways to think about NMR wipeout effects observed in cuprates, pnictides and other complex systems and their relation to quantum entanglement and fidelity in those systems.
}

\begin{figure*}[!h]
   \includegraphics[width=\linewidth]{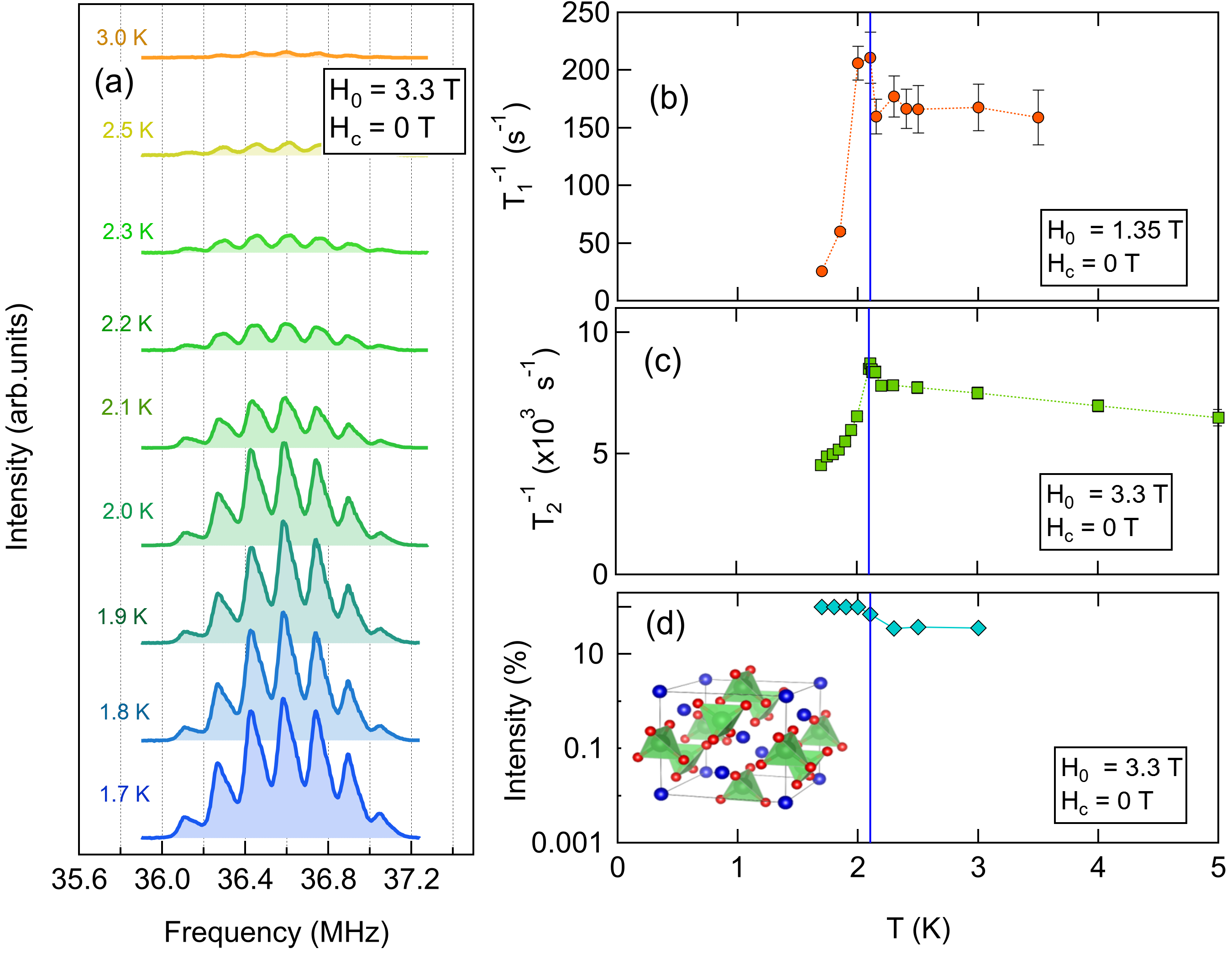}
    \caption{\label{fig:spectraVStemp} \textbf{NMR Spectra across $T_Q$.} (a) $^{51}$V NMR spectra for $H_c=0$ as a function of temperature. The spectra have been offset vertically for clarity. Panels (b,c,d) show the spin-lattice relaxation rate, \slrr, the decoherence rate $T_2^{-1}$, and the normalized intensity as a function of temperature.  All measurements were conducted with the applied field oriented perpendicular to (001). The vertical blue line indicates $T_Q$, and the inset shows the \tmvo\ unit cell, with Tm (blue), V (green), and O (red) atoms. }
\end{figure*}

\section*{Results}

Fig. \ref{fig:spectraVStemp}(a) shows $^{51}$V NMR spectra for a series of temperatures crossing the ferroquadrupolar transition at $H_c = 0$ (with the applied field in the $ab$ plane).  The spectra reveal seven peaks split by the nuclear quadrupolar interaction, but no discernable change in the overall shift or the quadrupolar splitting between the peaks.    Both the spin lattice relaxation rate, \slrr, and the spin echo decoherence rate, $T_2^{-1}$, shown in panels (b) and (c), exhibit small peaks at $T_Q$, but are suppressed in the ordered state. Moreover, the integrated intensity of the signal, shown in panel (d), shows little to no change across the phase transition. These results are consistent with previous measurements at low fields  \cite{Vinograd2022}.

\begin{figure*}[!h]
   \includegraphics[width=\linewidth]{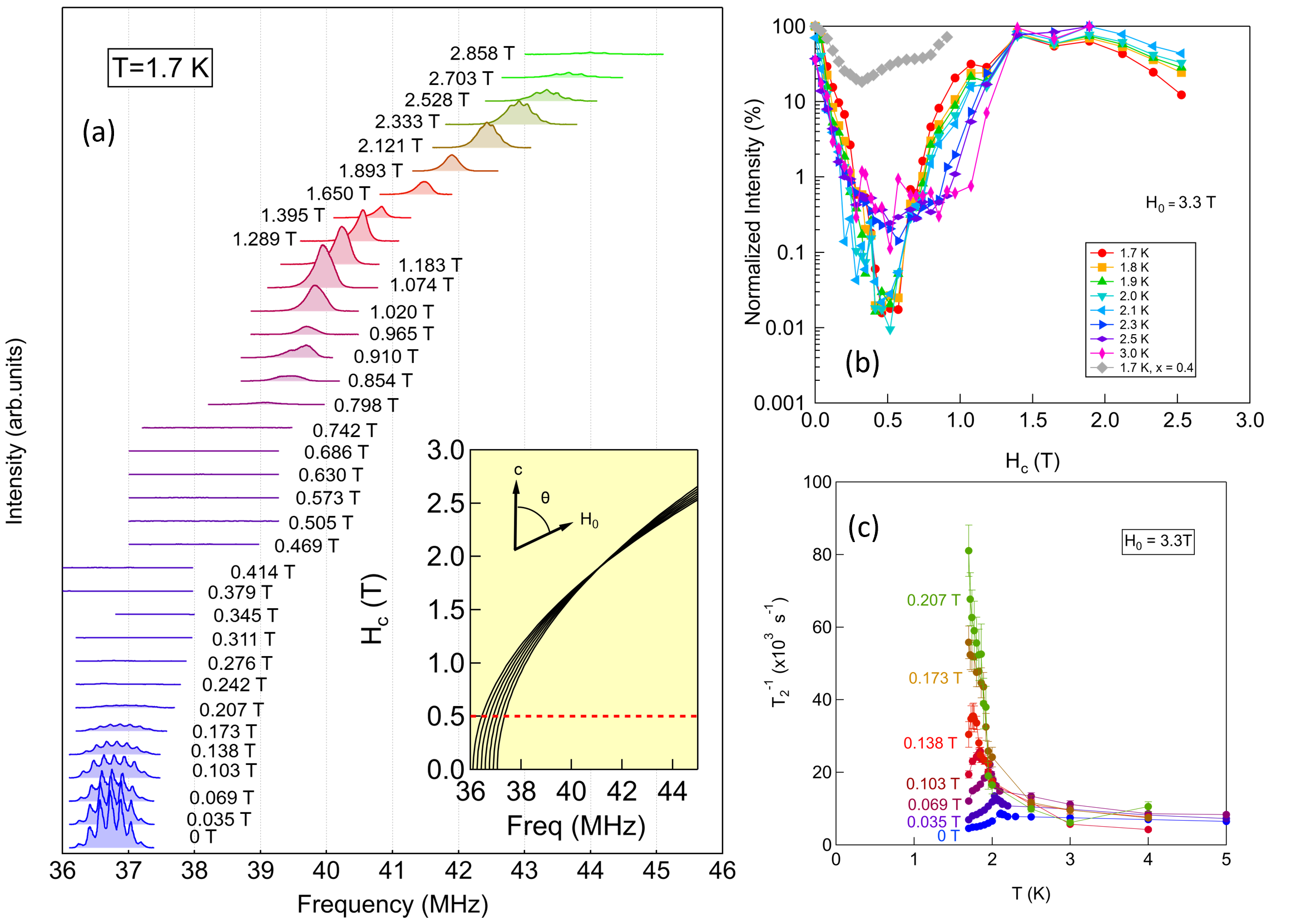}
    \caption{\label{fig:waterfall} \textbf{NMR Spectra in Transverse Fields.} (a) $^{51}$V NMR spectra at 1.7 K as a function of field along the $c$-axis, $H_c = H_0\cos\theta$ where $H_0 = 3.3$ T as depicted in the diagram.  The field is rotated in the (010) plane.   The spectra have been offset vertically for clarity, and have been normalized by the number of scans.  The inset depicts the calculated resonance frequencies of the seven satellites for different values of $H_c$.  The dashed red line indicates the critical field, $H_c^*$. (b) Integrated spin echo area times temperature normalized by the number of signal averages as a function of transverse field at various temperatures for $H_0 = 3.3$ T. The gray points correspond to Tm$_{0.6}$Y$_{0.4}$VO$_4$. (c) Decoherence rate, $T_2^{-1}$ versus temperature for several different values of $H_c$.}
\end{figure*}

The NMR response when crossing the phase transition as a function of $H_c$ at constant temperature is dramatically different.  Figure \ref{fig:waterfall} shows how the spectra evolve as the field $\mathbf{H}_0$ is rotated in the (010) plane at $T=1.7$ K.
Here, the $c$-axis projection, $H_c = {H}_0\cos\theta$, where $\theta$ is the angle relative to the $[001]$ direction.  The seven peaks in the spectrum shift upwards in frequency and closer together as $H_c$ increases. This behavior is well-described by the anisotropic Knight shift and electric field gradient tensors \cite{Wang2021}, as illustrated by the calculated curves in the inset. Surprisingly, the integrated intensity of the spectra is greatly reduced (wipeout) for a range of fields in the vicinity of $H_c^*=0.5$ T. This quantity is shown in panel (b) for a series of temperatures, and exhibits a drop of approximately four orders of magnitude near $H_c^*$.  The wipeout effect {is well known from the study of the cuprates} and arises due to a dramatic increase in the spin decoherence rate, $T_2^{-1}$ \cite{Curro2000,Curro2000b,ImaiCuprateWipeoutPRL,JulienGlassyStripes2001PRB,Ba122ClusterGlassNMR}. {An important difference, however, is that in \tmvo\ the effect occurs in a homogeneous system, without the presence of any dopants.  }  The spectra are obtained by measuring the size of the spin echo as a function of frequency for a fixed pulse spacing, $\tau$, and the echo size is proportional to $\exp[-2\tau/T_2]$. The intensity decreases, and because the spectrometer cannot operate for arbitrarily small $\tau$, the echo intensity will vanish for sufficiently large $T_2^{-1}$. This interpretation is supported by direct measurements of $T_2^{-1}$ for small $H_c$, shown in panel (c) for fields up to 0.2 T. Beyond this field it is not possible to obtain a direct measurement due to the wipeout effect.

\begin{figure}[!h]
    \includegraphics[width=\linewidth]{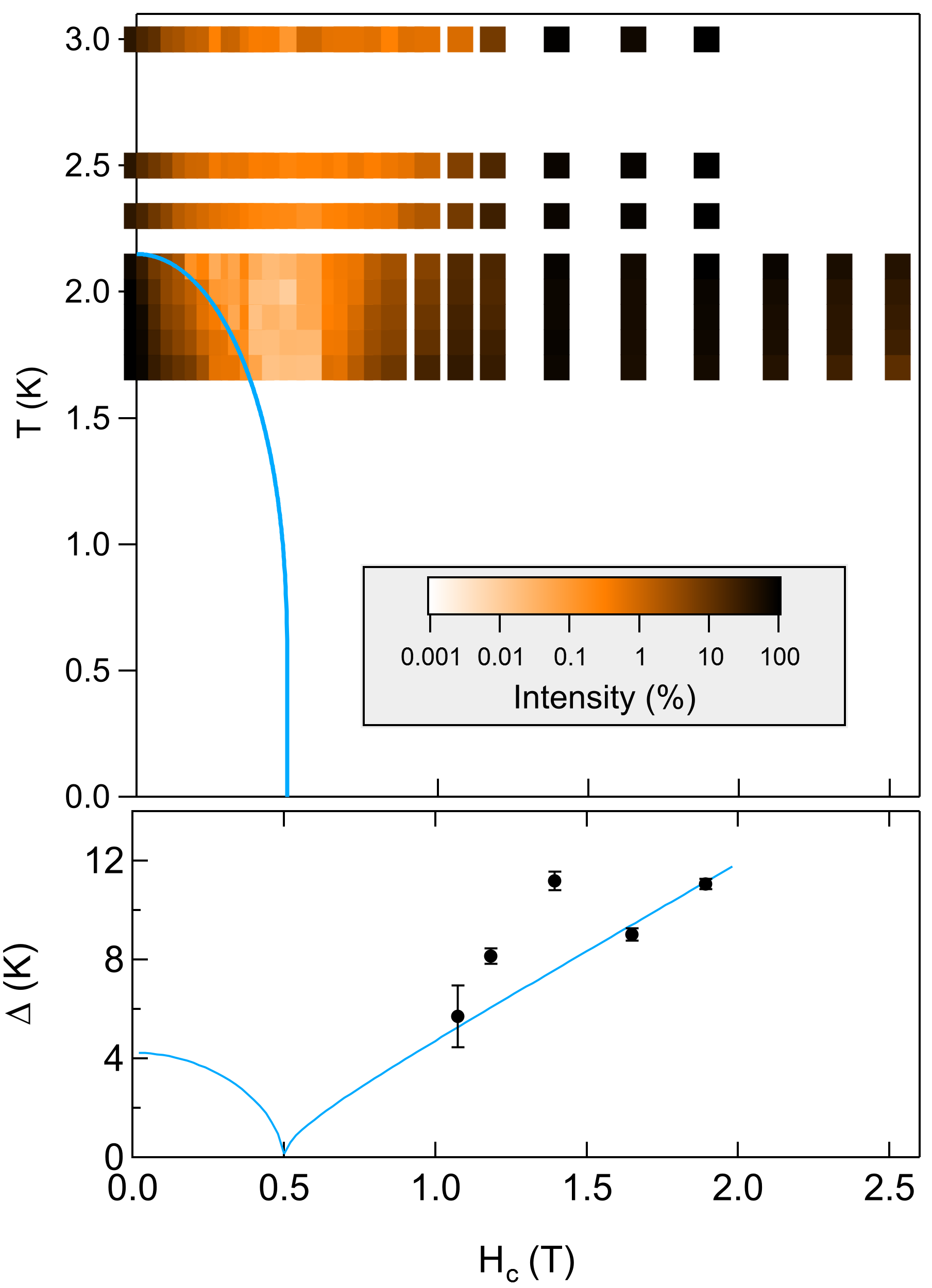}
    \caption{\label{fig:wipeout} \textbf{Quantum Critical Fan and Gap.} (a)  Phase diagram, with measured points colored by echo intensity.  The solid blue line is the ferroquadrupolar ordering temperature, $T_Q(H_c)$, reproduced from \cite{Massat2021}. (b) The calculated gap, $\Delta$ (solid line), and the measured gap ($\bullet$) extracted from fits to the \slrr, as described in the text. $\Delta(H_c=0)$ is set to 4.2 K as reported in \cite{Becker1972}. }
\end{figure}

This wipeout of the NMR signal persists to higher temperatures, as shown in Fig. \ref{fig:waterfall}(b).  In this case, the spectra are integrated as a function of frequency, and normalized by the value at 1.5 T at each temperature.  The wipeout effect in the vicinity of $H_c^*$ persists up to temperatures well above the ordering temperature $T_Q=2.15$ K.  Moreover, the range of fields throughout which the signal experiences wipeout broadens with temperature, giving rise to a fan-shaped region emerging from the quantum critical point, as illustrated in Fig. \ref{fig:wipeout}(b).

In order to check whether this wipeout is related to the QCP, we also measured the signal in Tm$_{0.6}$Y$_{0.4}$VO$_4$. Long range ferroquadrupolar order vanishes in the Y-doped material beyond a critical doping level of $\sim 0.2$. The intensity versus $H_c$ data for this compound, shown as gray points in Fig. \ref{fig:waterfall}(a), exhibits a slight reduction, but the effect is much less than that observed in the pure \tmvo.  These results point to quantum critical fluctuations as a mechanism for the signal wipeout in this material.

\begin{figure}[!h]
    \includegraphics[width=\linewidth]{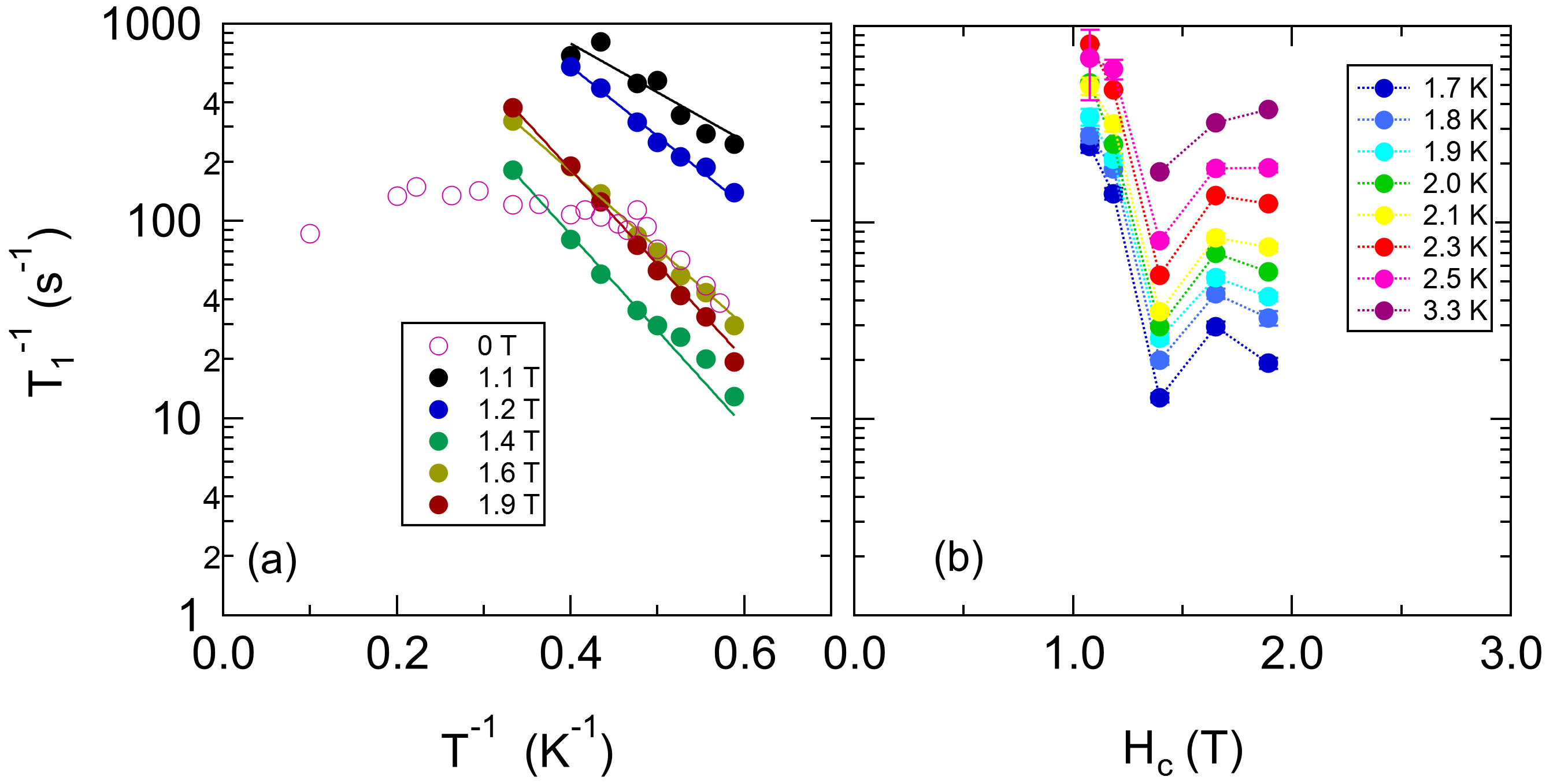}
    \caption{\label{fig:T1} \textbf{Spin Lattice Relaxation Rate.} The spin lattice relaxation rate of the $^{51}$V as a function of inverse temperature (a) and field (b). The solid lines are fits to an activated behavior, as discussed in the text. The open circles corresponding to $H_c = 0$ are reproduced from \cite{Vinograd2022} for the magnetic relaxation channel.}
\end{figure}

For sufficiently large $H_c$, the signal intensity recovers and it is possible to directly measure \slrr. Fig. \ref{fig:T1} shows how this quantity varies as a function of temperature and field. In this range, we find that \slrr\ exhibits activated behavior, $T_1^{-1} \propto \exp[-\Delta/T]$.  The gap, $\Delta$, is shown as a function of $H_c$ in Fig. \ref{fig:wipeout}(b).

\section*{Discussion}


The Tm ground state doublets are described by the  Hamiltonian:
\begin{equation}\label{eqn:TFIM}
\mathcal{H} = -J\sum_{i, j\in n.n.}\hat{\sigma}_{z}(i)\hat{\sigma}_{z}(j)  + \lambda \sum_{i} \hat{\sigma}_{x}(i)
\end{equation}
where $\sigma_z(i)$ is the $B_{2g}$ quadrupolar moment of the Tm at site $i$, $\sigma_x$ is the spin moment of the Tm along the $c$-axis, $J$ is the ferroquadrupolar exchange coupling, $\lambda = g_c\mu_B H_c$, and $g_c =10$ for \tmvo\ \cite{FisherNematicQCP}. The coupling to the nuclei is given by the hyperfine interaction:  $\mathcal{H}_{hyp}=   A_{cc}\sum_i I_{x}(i) \sigma_{x}(i)$, where $ A_{cc}/J \approx 2\times 10^{-4}$ and $I_x(i)$ is the nuclear spin along the $c$-axis (see Methods) \cite{Washimiya1970}. A similar problem was investigated for the case of a single central spin coupled to a 1D ring of Ising variables at $T=0$ in order to investigate the decoherence of the central spin state as a consequence of its coupling to a quantum critical environment \cite{Quan2006}.  In that case, a pure state of the central spin quickly entangles with the environment forming a mixed state, and the decoherence rate reaches a maximum at the QCP.  This model can be generalized to capture the relevant physics of the $^{51}$V NMR signal in \tmvo, in which there is a nuclear spin located at each lattice site (see Methods). The NMR free induction decay (FID) amplitude can be expressed as: $L_{FID}(t) =  |\langle\phi_g(t)|\phi_e(t)\rangle|$, where $|\phi_{e,g}(t)\rangle = \exp(-i\mathcal{H}_{e,g} t)|\phi(0)\rangle$ and $\mathcal{H}_{g,e} = \mathcal{H} \pm A_{cc}\sigma_x/2$. In other words, the FID is measure of the fidelity between two states that evolve from an initial state, $|\phi(0)\rangle$, under slightly different Hamiltonians in which the transverse field is either $H_c \pm A_{cc}/2g_c\mu_B$.   Note that if $\mathcal{H}_{hyp}$ includes terms coupling to $\sigma_z$, then the NMR response is no longer a simple function of the fidelity.  In a spin echo experiment, the nuclear spins are refocused with a $180^{\circ}$ pulse at time $\tau$, and the echo forms at time $2\tau$ with amplitude $L_{echo}(2\tau)= |\langle\phi_{ge}(2\tau)|\phi_{eg}(2\tau)\rangle|$ where $|\phi_{eg,ge}(2\tau)\rangle = \exp(-i\mathcal{H}_{g,e} \tau)\exp(-i\mathcal{H}_{e,g} \tau)|\phi(0)\rangle$.  If the system is far from the QCP then a slightly different transverse field will not significantly affect the time evolution of the state and the fidelity will decay slowly.  However, close to the QCP a small change of the transverse field can dramatically alter the wavefunction and the fidelity will rapidly decay with time.

At finite temperatures, the electronic wavefunction is no longer in a pure state and thermally excited states with different local fields will interact with the nuclei.  Both quantum fluctuations driven by the intrinsic dynamics of the system as well as incoherent thermal fluctuations will be contribute to the decoherence of the FID signal, washing out the enhancement at the QCP.  On the other hand, in a spin echo experiment the refocusing pulse dynamically decouples low frequency fluctuations  \cite{PulseSequenceFilterFunctionsPRB}. Chen et al. showed that the spin echo remains sensitive to the quantum critical fluctuations even at high temperatures in the 1D transverse field Ising model \cite{Chen2013}. Our results shown in Figs. \ref{fig:spectraVStemp}(b) and \ref{fig:wipeout} confirm this interpretation.  The spin echo intensity is suppressed by quantum critical fluctuations to temperatures above $T_Q$.

The NMR decoherence rate reflects an enhanced fidelity susceptibility, which can be related to the dynamical structure factor \cite{You2007}:
\begin{equation}
S_{xx}(\omega) = \int_{0}^{\infty}\langle \sigma_x(\tau)\sigma_x(0)\rangle e^{i\omega\tau}d\tau.
\end{equation}
The fluctuations $\langle\delta^2\sigma_x\rangle = \langle \sigma_x^2\rangle -\langle \sigma_x\rangle^2$ can be written as the sum of a thermal contribution $\langle\delta^2\sigma_x\rangle_T$ and a quantum contribution $\langle\delta^2\sigma_x\rangle_Q$ \cite{Frerot2016}. The incoherent thermal fluctuations are related to the transverse field susceptibility, $\chi_x$, via the classical fluctuation-dissipation theorem: $\langle\delta^2\sigma_x\rangle_T = \chi_x k_BT$.  These thermal fluctuations contribute primarily to the low frequency behavior of $S_{xx}(\omega)$. Coherent quantum fluctuations contribute to $S_{xx}(\omega)$ at finite frequency, {away from the quantum critical point}.  The decoherence of an FID can also be derived via  Bloch-Wangsness-Redfield theory, which  predicts an exponential decay $L_{FID}\sim e^{-t/T_2}$ where $T_2^{-1} = {A_{cc}^2}S_{xx}(0)/{2\hbar^2}$ \cite{Wangsness1953,Redfield1957,CPSbook}. For a spin echo, however, $L_{echo}(2\tau)$ can be expressed as a convolution of $S_{xx}(\omega)$ with a filter function, $F(\omega \tau)$ that encapsulates the influence of the $180^{\circ}$ refocusing pulse \cite{PulseSequenceFilterFunctionsPRB}.  In this case contributions of $S_{xx}(\omega)$ at frequencies $\omega\ll 1/\tau\approx 10^5$ Hz are filtered out.  As a result, contributions from thermal fluctuations are removed from the spin echo decoherence rate, but not quantum fluctuations at higher frequencies \cite{Chen2013}.  This distinction explains not only why the V signal intensity is able to map out the quantum critical fan but also why $T_2^{-1}$ changes little across the thermal phase transition at $H_c=0$ in Fig. \ref{fig:spectraVStemp}(c), where there are little to no quantum fluctuations.

{
In mean-field theory, the static transverse susceptibility decreases monotonically with field as $(\sech{(\lambda/T)})^2/T$ in the paramagnetic phase, which at low temperatures has the activated form $\frac{1}{T}\exp(-\Delta/T)$ consistent with the excitation gap $\Delta=2\lambda$.
In contrast, for the 3-dimensional TFIM \cite{Weihong1994} the $T=0$ transverse-susceptibility is known to diverge as one approaches the QCP from {\it either} side of the transition \cite{Brezin,Wegner}. Our quadrupolar system differs from a spin model in that the coupling of the order parameter to the lattice reduces the upper critical dimension to two and mean-field exponents are expected \cite{Schmalian,Paul,Qi-Xu} at the transition. Nevertheless, the gap must go to zero at the quantum critical point and the transverse susceptibility must have a maximum near QCP.
}

\begin{figure}[!h]
    \includegraphics[width=\linewidth]{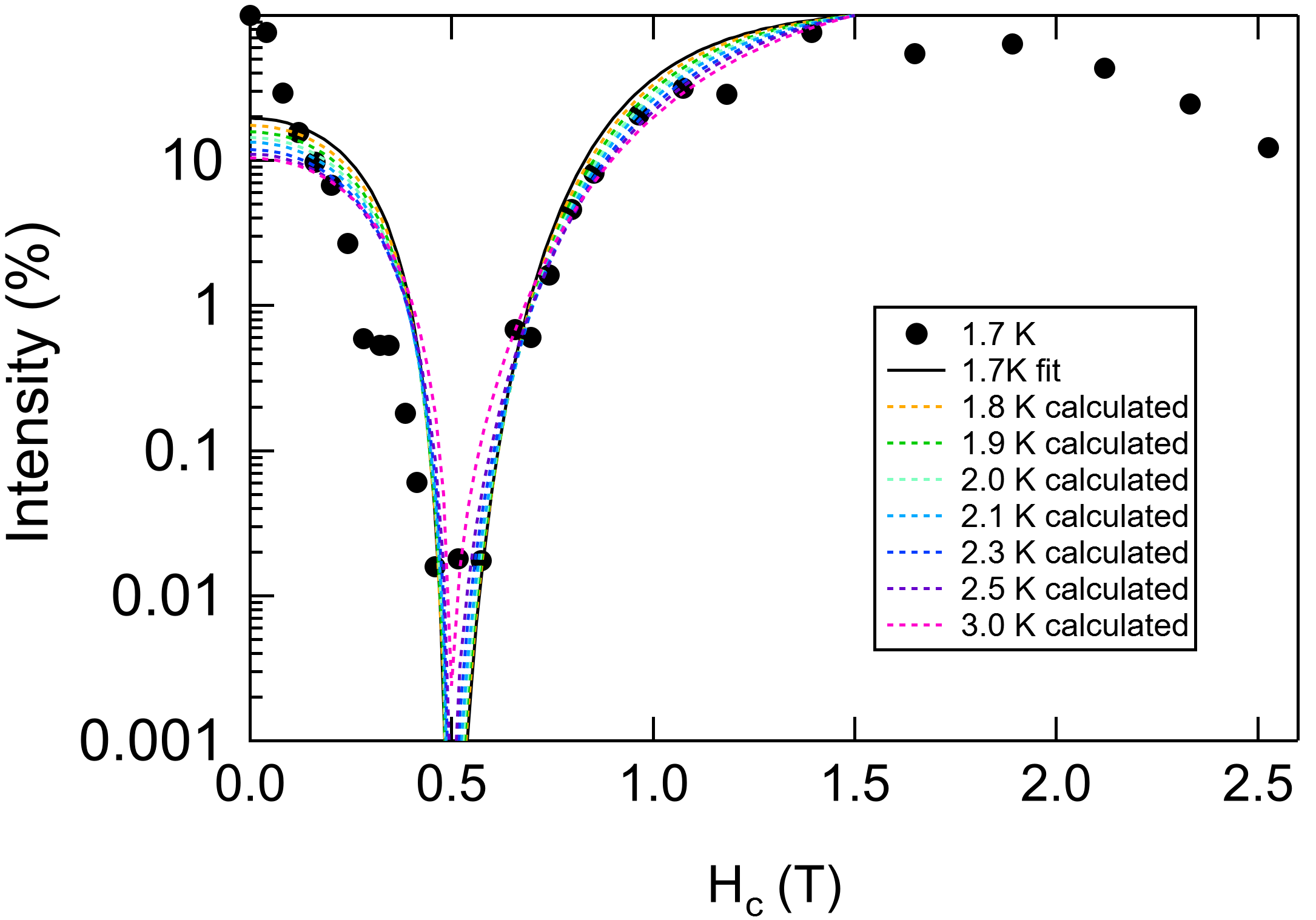}
    \caption{\label{fig:fits} \textbf{Echo Intensity Fits.} The normalized echo intensity ($\bullet$) at 1.7 K versus transverse field.  The solid black line is a fit to Eq. \ref{eqn:int} at 1.7 K with $\alpha = 36\pm 8$, and the dashed colored lines correspond to the same function computed at different temperatures.  }
\end{figure}

A recent theoretical study of the 1D TFIM also predicted that $S_{xx}(\omega)$  should exhibit thermally activated behavior at low frequencies: $S_{xx}\sim \exp[-\Delta/T]$ for $\omega\ll T\ll \Delta$, where $\Delta$ is the excitation gap in the quantum disordered regime \cite{Yang2022}.
We therefore postulate that the NMR intensity can be described {both in the disordered and the ordered states as}:
\begin{equation}\label{eqn:int}
  I(T,H_c) = I_0 \exp[-\alpha e^{-\Delta(H_c)/T}/T]
\end{equation}
where $\alpha \sim \tau A_{cc}^2$ is a constant.  The functional form of $\Delta(H_c)$ has been computed as a function of transverse field for different 3D cubic lattices (see Methods) \cite{Weihong1994}, and is constrained by the measured value of $\Delta(0) = 4.24$ K in the limit $H_c\ll H_c^*$ \cite{Becker1972}. For $H_c\gg H_c^*$,  $\Delta$ approaches $g_c\mu_B H_c$, the single ion gap for the ground state doublet \cite{Gehring1975,BleaneyTmVO4review}. This function is shown in Fig. \ref{fig:wipeout}(b).
We fit the intensity data at 1.7 K to Eq. \ref{eqn:int}, as shown in Fig. \ref{fig:fits}. The intensity is normalized by the value at $H_c = 1.5$ T for each temperature, and the only variable parameter is $\alpha$.  There is excellent agreement, and the increasing width as a function of temperature agrees well with the observed trend as shown in Figs. \ref{fig:waterfall}(b) and \ref{fig:wipeout}(a). The fit is less good for low $H_c$, and we speculate that this behavior may reflect the fact that there remains a finite field $H_0\sin\theta$ in the longitudinal direction in this limit.  The Ising variables couple to this field to second order \cite{Vinograd2022}, which may give rise to higher order effects not captured by the fitting function.

The gap extracted from \slrr\ measurements compares well with the expected values of $\Delta(H_c)$ as shown in Fig. \ref{fig:wipeout}(b), however it has been argued that $T_1^{-1}\sim \exp[-2\Delta/T]$ \cite{Yang2022}.  This result was based on the assumption of a hyperfine interaction that includes a coupling to $S_{zz}(\omega)$, the structure function for longitudinal fields.  Such a coupling is absent in \tmvo, but spin lattice relaxation is driven by fluctuations perpendicular to the quantization axis, $\mathbf{H}_0$, which is not parallel to the transverse field in our case.  Therefore, $S_{xx}$ can contribute to \slrr, and hence we expect $T_1^{-1}\sim \exp[-\Delta/T]$.

It is noteworthy that both TmVO$_4$ and LiHoF$_4$ are model systems for the 3D TFIM due to the non-Kramer's nature of the ground state $4f$ electrons, but an important difference between the two, aside from the nature of the long-range order, is the anisotropy of the hyperfine coupling to the nuclear spin degrees of freedom.  In the latter, the $^{165}$Ho nuclear spin couples to the 4f magnetic moment along the longitudinal (easy) axis, giving rise to a composite spin at low temperatures which modifies the shape of the phase boundary near the QCP \cite{RosenbaumQCmodel}.   Consequently a larger transverse field is required in order to fully destroy the ferromagnetic state. In \tmvo, the $^{169}$Tm nuclear spin also couples to the 4f magnetic moment, but only along the transverse direction \cite{BleaneyTmVO4review}.


Our observation of NMR wipeout of the $^{51}$V in \tmvo\ provides experimental confirmation that a qubit coupled to a quantum critical environment quickly transitions from a pure to a mixed state \cite{Quan2006}. It is perhaps natural to expect that such an environment would be detrimental to any application in which the qubits are used as a resource for quantum information.  Conversely, these results also suggest that an ensemble of strongly coupled qubits, such as in a spin-based quantum computer, would similarly be sensitive to noise arising from nearby nuclear spins which could limit the coherence time for quantum computations.

\section*{Methods}

\subsection*{Sample Preparation}

Single crystals of \tmvo\ were grown from a Pb$_2$V$_2$O$_7$ flux using 4 mole percent of Tm$_2$O$_3$, following the methods described in \cite{FEIGELSON1968,Smith1974}. The sample was secured with superglue inside the coil to prevent motion due to torque arising from the anisotropic susceptibility. The coil was attached to a platform on the (100) face, that was in turn inserted in a dual-axis goniometer.  The crystal was rotated in the (010) plane, and the orientation was controlled with a precision $<0.1^{\circ}$ \cite{Wang2021}. The orientation $\mathbf{H}_0 \parallel [100]$  was identified by minimizing \slrr.  The crystal was rotated in the (010) plane, which is crucial because the second order Zeeman interaction broadens the phase transition for in-plane field components along the [110] directions \cite{Vinograd2022}.   The applied field ${H}_0 = 3.3$ T is well below the in-plane critical field $\sim 9$ T identified previously  \cite{Vinograd2022}, so the in-plane field component has a negligible effect on the Tm $4f$ degrees of freedom.

\subsection*{Spectral Structure}

$^{51}$V has spin $I=7/2$, with seven transitions at resonance frequencies given by
\begin{equation}
\nu_n = \gamma H_0 (1 + K_{aa}\sin^2\theta + K_{cc}\cos^2\theta) + n\frac{\nu_{cc}}{2}(3\cos^2\theta-1),
\end{equation}
where $\gamma = 11.193$ MHz/T is the gyromagnetic ratio, $K_{\alpha\alpha}$ and $\nu_{\alpha\alpha}$ are the Knight shift and EFG tensors, and  $n = -3, \cdots ,+3$.  These frequencies are shown in the inset of Fig. \ref{fig:waterfall} using the parameters $H_0 = 3.3$ T, $K_{aa}=-0.96$\%, $K_{cc}=36$\%, $\nu_{cc}=0.62$ MHz \cite{Wang2021}.  The spectra shown in Fig. \ref{fig:waterfall} were measured by a spin echo pulse sequence: $\frac{\pi}{2} - \tau - \pi - \tau - echo$ is applied at a frequency, $f$. The area under the spin echo signal is plotted as a function of excitation frequency, for a fixed value of the pulse spacing $\tau$.

\subsection*{Spin Lattice Relaxation Measurements}

\slrr\ was measured by inversion recovery for sufficiently high fields $H_{c} > 1$ T where the spectral intensity is recovered.  Note that previous studies identified significant contributions to the spin lattice relaxation from quadrupolar fluctuations, giving rise to complicated relaxation behavior at each transition \cite{Vinograd2022}. However, in this regime of fields the seven satellites overlap, giving rise to a single feature in the spectra, as shown in Fig. \ref{fig:waterfall}, precluding a detailed analysis. The spin lattice relaxation rate, \slrr, was measured by inversion recovery at the central transition ($I_z = +1/2 \leftrightarrow -1/2$). The magnetization recovery was fit to the standard expression for a spin 7/2 nucleus:
$M(t)\!=\!M_0\left(1-2f\sum_n A_n e^{-\alpha_n t/T_1}\right)$
where $M_0$ is the equilibrium nuclear magnetization, $f$ is the inversion fraction, $A_1 = 1225/1716$, $A_2 = 75/364$, $A_3 = 3/44$, $A_4= 1/84$, $\alpha_1 = 28$, $\alpha_2 = 15$, $\alpha_3  = 6$, and $\alpha_4 = 1$. The data in Fig. \ref{fig:T1}(b) reveal a sharp dip in the magnitude of \slrr\ for $H_c\approx 1.5$ T. The origin of this change is not known, but we speculate that this effect may be an artifact of the overlapping transitions near $H_c = H_0\cos\theta_m$, where $\theta_m \approx 54.7^{\circ}$, because in this case the recovery function will change.

\subsection*{Hyperfine Coupling}

The coupling between a nuclear spin and a non-Kramer's doublet in a magnetic field is given by \cite{Washimiya1970}:
\begin{eqnarray}
\label{eqn:hyp2}
\nonumber \mathcal{H}_{hyp}&=&   C(H_z I_z - H_y I_y)\sigma_z+ C(H_zI_y + H_yI_z)\sigma_y \\
&&+ A_{cc}I_x \sigma_x.
\end{eqnarray}
The first two terms are hyperfine couplings to induced moments, and the coupling constants are given by: $C  = \frac{1}{2}\gamma\hbar (g_J \mu_B)^2 b ({h_a}/{\mu_a})$ and $A_{cc} = \gamma\hbar g_c \mu_B ({h_c}/{\mu_c})$, where $g_J = 7/6$ and $b/k_B = 0.082 K^{-1}$ \cite{BleaneyTmVO4review}. The ratios $(h_{\alpha}/{\mu_{\alpha}})$ can be determined by calculating the direct dipole fields at the V sites in the lattice: $h_{a}/{\mu_{a}} = -0.0336 T/\mu_B$ and $h_{c}/{\mu_{c}} = 0.0671 T/\mu_B$. We find that $C = -0.37 \mu$K/T, $A_{cc} = 368 \mu$K, and the ratio $C H_0/A_{cc} \sim 3\times 10^{-3}$ for $H_0 = 3.3$ T, thus we can safely neglect the first two terms in Eq. \ref{eqn:hyp2}. Note that each V is coupled to multiple Tm sites, but the dominant ones are the two nearest neighboring Tm above and below each V along the $c$-axis.  For simplicity, we assume that each V couples to a single Tm, and we assume $I=1/2$.

\subsection*{Generalization to Nuclei in a Lattice}
We consider a generalization of Eq. \ref{eqn:TFIM} to include a nuclear spin at each lattice site: $\mathcal{H}+ \mathcal{H}_n + \mathcal{H}_{hyp}$, where $\mathcal{H}$ is defined in Eq. \ref{eqn:TFIM},
\begin{equation}
\mathcal{H}_n =\frac{1}{2}\omega_L \sum_{j=1}^N  (|e_j\rangle\langle e_j| -|g_j\rangle\langle g_j| )
\end{equation}
describes the Zeeman interaction of the spins (which we are considering to be spin 1/2 qubits instead of spin 7/2), and
\begin{equation}
\mathcal{H}_{hyp} = \frac{1}{2}A_{cc} \sum_{j=1}^N  (|e_j\rangle\langle e_j| -|g_j\rangle\langle g_j| )\sigma_x(j)
\end{equation}
describes the hyperfine coupling. Here $\omega_L = \gamma H_c$ is the Larmor frequency, and $|g_j\rangle$ and $|e_j\rangle$ describe the eigenstates of the $I_x(j)$ operator at site $j$.  Note that this interaction describes coupling to nuclear spins at each site $j$, in contrast to that in \cite{Quan2006} and \cite{Chen2013},  with a single qubit coupled to {all} sites.

In an NMR experiment, all nuclear spins are initialized in the fully polarized state, such that the total wavefunction is $\frac{1}{\sqrt{2}} |\psi_g\rangle + \frac{1}{\sqrt{2}} |\psi_e\rangle $, where $|\psi_g\rangle = {|g_1\rangle \otimes\cdots \otimes|g_N\rangle\otimes|\phi(0)\rangle}$ and $|\psi_e\rangle = {|e_1\rangle \otimes\cdots \otimes|e_N\rangle\otimes|\phi(0)\rangle}$, where $|\phi\rangle$ is the many-body wavefunction of the Tm system. Under the action of $\mathcal{H}$ the wavefuction will evolve to:
\begin{eqnarray}\label{}
  \nonumber |\psi(t)\rangle &=& \frac{1}{\sqrt{2}}e^{i\omega_L t/2} |g_1\rangle \otimes\cdots \otimes|g_N\rangle\otimes e^{-i\mathcal{H}_g t}|\phi_g(t)\rangle+\\
  \nonumber &&\frac{1}{\sqrt{2}}e^{-i\omega_L t/2} |e_1\rangle \otimes\cdots \otimes|e_N\rangle\otimes e^{-i\mathcal{H}_e t}|\phi_e(t)\rangle,
\end{eqnarray}
where $\mathcal{H}_{g,e} = \mathcal{H} \pm A_{cc}\sigma_x/2$.  The free induction decay (FID) is then given by:
\begin{equation}\label{eqn:FID}
  L_{FID}(t) =\sqrt{\langle I_z\rangle ^2 + \langle I_y\rangle^2}/(N/2) = |\langle \phi_g(t)|\phi_e(t)\rangle|.
\end{equation}
 {(Note that in \cite{Quan2006} the Loschmidt echo is defined as the square of this quantity.)} In a spin echo experiment, a $180^{\circ}$ pulse swaps the $|e_j\rangle$ and $|g_j\rangle$ states, and the echo forms at time $2\tau$.  In this case the magnitude of the echo is given by:
\begin{equation}\label{eqn:echo}
  L_{echo}(2\tau) =|\langle \phi_{ge}(2\tau)|\phi_{eg}(2\tau)\rangle|,
\end{equation}
where $|\phi_{eg,ge}(2\tau)\rangle = e^{i\mathcal{H}_{g,e}\tau}e^{i\mathcal{H}_{e,g} \tau}|\phi(0)\rangle$.

These expressions can be re-written in terms of the density matrix, $\rho$, of the Tm system which is appropriate for finite temperatures.  In this case we have:
\begin{eqnarray}
  L_{FID}(t) &=& \left|Tr\left\{ e^{i\mathcal{H}_e t} \rho e^{-i\mathcal{H}_g t} \right\}  \right| \\
  L_{echo}(2t) &=& \left|Tr\left\{ e^{i\mathcal{H}_gt}e^{i\mathcal{H}_et} \rho e^{-i\mathcal{H}_gt}e^{-i\mathcal{H}_et} \right\}  \right|.
\end{eqnarray}

\subsection*{Excitation gap for 3D lattice}

For a 3D cubic lattice, the excitation gap for the transverse field Ising model varies as:
\begin{equation}
  \Delta =
  \begin{cases}
g_c\mu_B H \sqrt{1-x}\left[(-\ln(1-x)/x\right]^{-1/6}  & x< 1\\
\Delta(0)\sqrt{1-x^{-2}}\left[(-\ln(1-x^{-2})/x^{-2}\right]^{-1/6} & x >1
  \end{cases}
\end{equation}
where $x=H_c^*/H_c$.\\

\section*{Data Availability Statement}

All data needed to evaluate the conclusions are present in the paper and/or supplemental materials. Correspondence and requests for materials should be addressed to N.J.C.

\section*{Acknowledgements}

We acknowledge helpful discussions with A. Albrecht, R. Baunach and F. Ronning. Work at UC Davis was supported by the NSF under Grants No. DMR-1807889 and PHY-1852581, as well as the UC Laboratory Fees Research Program ID LFR-20-653926. Crystal growth performed at Stanford University was supported by the Air Force Office of Scientific Research under award number FA9550-20-1-0252. P. M. was partially supported by the Gordon and Betty Moore Foundation Emergent Phenomena in Quantum Systems Initiative through Grant GBMF9068.

\section*{Competing interest}

The authors declare no competing financial or non-financial interests.

\section*{Author Contributions}

N.J.C. planned the project; P. M., M.Z. and I.R.F. synthesized, characterized, and prepared the single crystals; Y.-H. N. performed the NMR measurements with assistance from I. V., T. G. and C. C.; R. S., I.R.F., and N.J.C. wrote the manuscript with input from the other authors. All authors contributed to discussions of the results and interpretation.

\section*{References}
\bibliography{TMVO4QCpaper}

\end{document}